\documentclass{iopart}

\usepackage{graphicx}
\usepackage{dcolumn}
\usepackage{bm}
\usepackage{color}
\usepackage{iopams}

\newcommand{\expect}[1]{\langle #1 \rangle}
\newcommand{\eps}{\varepsilon}
\newcommand{\w}{\omega}
\newcommand{\s}{\sigma}
\newcommand{\G}{\Gamma}
\newcommand{\up}{\uparrow}
\newcommand{\down}{\downarrow}

\newcommand{\beq}{\begin{eqnarray}}
\newcommand{\eeq}{\end{eqnarray}}
\newcommand{\be}{\begin{equation}}
\newcommand{\ee}{\end{equation}}

\begin{document}

\title[Asymmetry-induced effects in Kondo quantum dots with FM leads]
{Asymmetry-induced effects in Kondo quantum dots coupled to ferromagnetic leads}

\author{K. P. W\'ojcik$^1$, I. Weymann$^1$, J. Barna\'{s}$^{1,2}$}

\address{$^1$ Faculty of Physics,
Adam Mickiewicz University, 61-614 Pozna\'n, Poland}
\address{$^2$ Institute of Molecular Physics, Polish Academy of
Sciences, 60-179 Pozna\'{n}, Poland}
\ead{weymann@amu.edu.pl}

\date{\today}

\begin{abstract}
We study the spin-resolved transport through single-level quantum dots
strongly coupled to ferromagnetic leads in the Kondo regime,
with a focus on contact and material asymmetry-related effects.
By using the numerical renormalization group method, we analyze the dependence of
relevant spectral functions, linear conductance and tunnel magnetoresistance
on the system asymmetry parameters. In the parallel magnetic configuration of the device
the Kondo effect is generally suppressed due to the presence of exchange field, irrespective of
system's asymmetry. In the antiparallel configuration, on the other hand, the Kondo effect can develop
if the system is symmetric. We show that even relatively weak asymmetry
may lead to the suppression of the Kondo resonance in the antiparallel configuration
and thus give rise to nontrivial behavior of the tunnel magnetoresistance.
In addition, by using the second-order perturbation theory we derive general formulas
for the exchange field in both magnetic configurations of the system.
\end{abstract}

\pacs{72.25.Mk, 73.63.Kv, 85.75.-d, 73.23.Hk, 72.15.Qm}

\maketitle

\section{Introduction}

Transport properties of nanoscopic objects, such as quantum dots or molecules,
tunnel-coupled to external leads have attracted a lot of attention
from both theoretical as well as experimental sides~\cite{loss02,zutic04,maekawa06,seneor07,barnas08,bader10}.
This is mainly due to possible applications in nanoelectronics and spintronics,
and due to a unique possibility to study various many-body correlation effects
between single charges and spins.
When the coupling between the quantum dot and external leads is relatively strong,
the electronic correlations may lead to the Kondo effect if the dot's
occupation number is odd~\cite{kondo64,hewson_book,goldhaber-gordon_98,cronenwett_98}.
For quantum dots coupled to ferromagnetic leads, on the other hand, 
it was shown that the Kondo resonance can be suppressed due to the
presence of an effective exchange field, $\Delta\eps_{\rm exch}$, that 
leads to the spin splitting of the quantum dot level~\cite{martinekPRL03,martinekPRL03_2,choiPRL04,martinekPRB05,
pasupathy_04,heerschePRL06,hamayaPRB08,hauptmannNatPhys08,gaassPRL11}.
This suppression occurs if the magnetic moments of external leads
form a parallel (P) magnetic configuration and 
when $| \Delta \eps_{\rm exch}^{\rm P}| \gtrsim T_K$, where $T_K$ is the Kondo temperature
and $\Delta\eps_{\rm exch}^{\rm P}$ denotes the level splitting due to exchange field in the
parallel configuration~\cite{pasupathy_04}.
For an antiparallel magnetic configuration of the device, the exchange field was found to vanish,
$\Delta\eps_{\rm exch}^{\rm AP}\to 0$, since the effective coupling to external leads
becomes then spin-independent~\cite{martinekPRL03_2,choiPRL04,pasupathy_04,weymannPRB11}.
This is however true only for fully symmetric systems,
while, as we show in this paper,
for systems exhibiting some asymmetry, either the left-right contact asymmetry or the material's asymmetry,
the exchange field may also develop in the antiparallel configuration.
Since experimentally it is very difficult to build a truly symmetric device, 
it seems desirable to analyze the effects of contact and material's asymmetry
on spin-resolved transport properties of quantum dots.

In the present paper we thus thoroughly study transport through 
quantum dots coupled to ferromagnetic leads in the Kondo regime,
focusing especially on asymmetry-induced effects.
To obtain the correct picture, we employ the Wilson's numerical renomalization group (NRG)
method~\cite{WilsonRMP75,KrishnaMurphy80,BullaRMP08}
with the idea of a full density-matrix (fDM)~\cite{WeichselbaumPRL07}.
By using NRG, we calculate the dependence of relevant
spin-resolved spectral functions, the linear conductance
in the parallel and antiparallel configurations, and the tunnel magnetoresistance (TMR) on the
asymmetry between the couplings to the left and right leads and for different 
spin polarizations of the electrodes.
We show that even relatively small asymmetry
may fully suppress the Kondo resonance in the antiparallel configuration,
leading to nontrivial dependence of the TMR effect on the asymmetry parameters.
We also show that although asymmetry generally destroys the Kondo effect in the antiparallel configuration,
there is a range of asymmetry parameters, when the Kondo resonance can be restored.
In addition, by using the second-order perturbation theory, we derive general formulas
for the exchange field in both magnetic configurations depending on asymmetry parameters.

\section{Theoretical description}

\begin{figure}[t]
  \includegraphics[width=0.65\columnwidth]{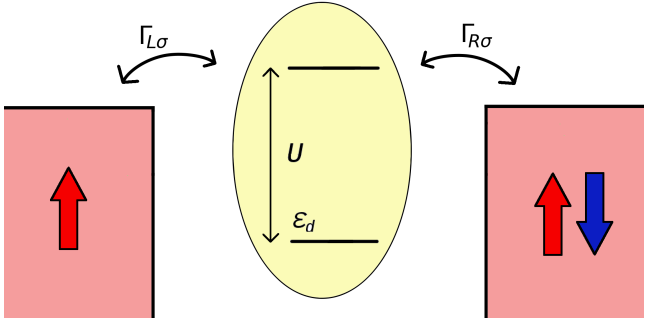}
  \caption{\label{Fig:1}
  (Color online)
  The schematic of a single-level quantum
  dot coupled to the left and right ferromagnetic electrodes.
  The magnetizations of electrodes are assumed to be collinear
  and they can form either parallel (P) or antiparallel (AP)
  magnetic configuration, as indicated in the figure.
  The dot is coupled to the left and right lead
  with the coupling strength $\G_{r\s}$ ($r=L,R$). The dot level
  has energy $\eps_d$ and $U$ denotes Coulomb correlation on the dot.}
\end{figure}

The system consists of two ferromagnetic leads coupled to a single-level 
quantum dot, see Fig.~\ref{Fig:1}. The magnetizations of the leads
are assumed to be collinear and they can form two magnetic configurations:
the parallel (P) and antiparallel (AP) ones.
Switching between different magnetic configurations of the device
can be obtained by sweeping the hysteresis loop,
provided the left and right ferromagnets have different coercive fields.
The system considered can be described by the single-impurity Anderson Hamiltonian 
\beq
H &=& \sum_{r=L,R}\sum_{{\bf k}\s} \eps_{r {\bf k}\s} 
	a_{r {\bf k}\s}^\dagger a_{r {\bf k}\s}
	+ \eps_{d} \sum_{\s} n_\s + U n_\up n_\down 
	\nonumber\\
	&& + \sum_{r=L,R}\sum_{{\bf k}\s} \left( V_{r {\bf k}\s} 
	a_{r {\bf k}\s}^\dagger d_{\s} + h.c.\right),
\label{H}
\eeq
where $a_{r {\bf k}\s}^\dagger$ creates an electron with spin $\sigma$ momentum ${\bf k}$
and energy $\eps_{r {\bf k}\s}$ in the left ($r=L$) or right ($r=R$) lead.
The energy of the dot level is denoted by 
$\eps_d$, $U$ describes the Coulomb interaction of two electrons residing on the dot, and
$n_\s = d_{\s}^\dagger d_{\s}$ is the particle number operator,
with $d_{\s}^\dagger$ creating a spin-$\sigma$ electron on the dot.
The last term of the Hamiltonian (\ref{H}) describes tunneling processes
between the dot and the ferromagnetic electrodes,
with $V_{r {\bf k}\s}$ being the relevant hopping matrix elements.
Due to the coupling to external leads, the dot level acquires certain width,
which is described by the spin-dependent hybridization function, $\Gamma_{r\s} = \pi\rho V_{r\s}^2$, for 
lead $r$ and spin $\sigma$. Here, $\rho$ is the density of states at the Fermi level
and we assumed that $V_{r{\bf k}\s}$ is
independent of momentum $\bf{k}$~\cite{martinekPRL03_2,choiPRL04}.
The conduction bands of the leads are assumed to be energy and spin independent, $\rho = 1/(2W)$,
where $W\equiv 1$ is used as energy unit. 
These assumptions are justifiable since we are interested in the Kondo regime which
is mainly related with the conduction electron states around the Fermi energy.
Next, by introducing the spin polarization of lead $r$,
$p_r = (\G_{r\up}-\G_{r\down})/(\G_{r\up}+\G_{r\down})$,
the couplings can be expressed as, $\Gamma_{r\s} = (1+\s p_r)\Gamma_r$, with
$\Gamma_r = (\Gamma_{r\up}+\Gamma_{r\down})/2$.
Here, $\Gamma_{r\s}$ denotes the coupling to the spin-majority ($\s=\up$)
or spin-minority ($\s=\down$) conduction band of the ferromagnetic lead $r$.
For convenience, we have incorporated the effect of leads' ferromagnetism
into the spin-dependent tunnel matrix elements.
This assumption is commonly used in NRG calculations~\cite{martinekPRL03_2,weymannPRB11}.

The asymmetry induced effects to be studied in the present paper
will include the left-right contact asymmetry, $\G_L\neq\G_R$, and the material asymmetry,
i.e. different spin polarizations of the leads, $p_L\neq p_R$.
In particular, we will study how the linear response transport properties depend
on the parameters of the system, especially on the couplings' ratio $\G_{R} / \G_{L}$
and spin polarizations' ratio $p_{ R} / p_{L}$.
The linear-response spin-dependent conductance of the system
can be found from the Meir-Wingreen formula~\cite{meir},
\be\label{Eq:Glin}
  G_\s = \frac{e^2}{h} \frac{4\G_{L\s} \G_{R\s}} {\G_{L\s} + \G_{R\s}}
  \int \!\! d\w \left(-\frac{\partial f(\w)}{\partial\w}\right) \! \pi A_\s (\w)\,,
\ee
where $f(\w)$ denotes the Fermi function and $A_\s(\w)$ is the spectral function
of dot level. The spectral function is given by $A_\s(\w) = -\frac{1}{\pi} \Im{\rm m}\;G^{\rm R}_\s(\w)$,
where $G^{\rm R}_\s(\w)$ is the Fourier transform of the retarded d-level
Green's function, $G_\s^{\rm R}(t) = -i\theta(t) \expect{\{ d_\s(t),d_\s^\dag(0) \}}$.

Another quantity of interest is the tunnel magnetoresistance (TMR),
which describes the change in spin-dependent transport properties
when the magnetic configuration of the device is varied. The TMR is defined as~\cite{julliere},
\be \label{Eq:TMR}
  {\rm TMR} = \frac{G^{\rm P} -G^{\rm AP}  } {G^{\rm AP} } \,,
\ee
where $G^{\rm P/AP}$ is the total linear conductance in the P/AP magnetic configuration.
In addition, we will also study the spin filtering properties of the system,
which can be related to the spin polarization of the linear conductance,
\be \label{Eq:P}
  \mathcal{P}^{\rm P/AP} = \frac{G^{\rm P/AP}_\up - G^{\rm P/AP}_\down} {G^{\rm P/AP}_\up + G^{\rm P/AP}_\down} \,,
\ee
where $\mathcal{P}^{\rm P/AP}$ denotes the spin polarization in 
the parallel or antiparallel magnetic configuration of the device.

For the single-impurity Anderson model, it is very convenient to perform 
an orthogonal transformation from the left-right basis into an even-odd basis, where
only the even linear combination of the lead operators couples to the dot, while the odd
combination is completely decoupled. The new effective spin-resolved couplings
for the parallel and antiparallel configuration can be then expressed as
\be 	\label{Eq:Gammaeff}
  \G_\s^{\rm P/AP} = \G \left( 1 + \s p \beta^{\rm P/AP} \right) , 
\ee
where $p$ denotes the average spin polarization $p = (p_{ L} + p_{R}) / 2$
and $\G = \G_{L} + \G_{R}$, while the parameter $\beta^{\rm P/AP}$ is given by
\beq
\beta^{\rm P} &=& 1 + {(\G_{ L} - \G_{ R})(p_{ L} - p_{ R}) \over (\G_{ L} + \G_{ R})(p_{ L} + p_{ R})} ,
	\label{Eq:bP}\\
\beta^{\rm AP} &=& {(\G_{L} - \G_{ R}) \over (\G_{ L} + \G_{ R})} + {(p_{ L} - p_{ R}) \over (p_{ L} + p_{ R})} .
	\label{Eq:bAP}
\eeq
From the above formula follows that the couplings are generally different for each spin direction,
$\Gamma_\up^{\rm P/AP} \neq \Gamma_\down^{\rm P/AP}$,
which results in different level renormalization for spin-up and spin-down.
This difference leads to an effective spin-splitting of the dot level,
known as contact-induced exchange field, $\Delta\eps_{\rm exch}^{\rm P/AP}$
~\cite{martinekPRL03_2,choiPRL04,martinekPRB05}.
In the parallel configuration, $\Gamma_\up^{\rm P} \neq \Gamma_\down^{\rm P}$,
and consequently, $\Delta\eps_{\rm exch}^{\rm P} \neq 0 $, once
the spin polarization is finite, $p_r>0$, for any value of left-right contact asymmetry.
In the antiparallel configuration, on the other hand, 
the exchange field can occur, $\Delta\eps_{\rm exch}^{\rm AP} \neq 0 $, provided there is an asymmetry in the system, so that
$\Gamma_\up^{\rm AP} \neq \Gamma_\down^{\rm AP}$.
The asymmetry can be either related to the contacts, $\G_L \neq \G_R$,
or to the material, $p_L \neq p_R$. Moreover, it turns out that even in the presence of asymmetry,
in the antiparallel configuration there is a parameter range where
the exchange field can still vanish, which happens for $\beta^{\rm AP} = 0$.
This occurs precisely when the following condition is met
\be \label{Eq:exchvan}
 \frac{ \G_L - \G_R} {\G_L + \G_R} = - \frac{p_L - p_R}{p_L + p_R}\,.
\ee 

The spin-dependence of the effective couplings gives rise to the exchange field $\Delta\eps_{\rm exch}$,
which can be found analytically by calculating second-order
corrections $\delta\eps_{d\s}$ to the energy of quantum dot levels,
$\Delta\eps_{\rm exch}\equiv \delta\eps_{d\up} - \delta\eps_{d\down}$.
At low temperature, one then gets for the parallel and antiparallel magnetic configurations
\be \label{Eq:Exch}
  \Delta\eps_{\rm exch} ^ {\rm P/AP} = \frac{2}{\pi} \beta ^{\rm P/AP} p\G \ln \left| \frac{\eps_d}{\eps_d+U}\right| \,.
\ee
There are actually two factors that determine the strength of the exchange field:
the first one, $\sim \ln \left| \eps_d/(\eps_d+U)\right|$, is related to the gate voltage,
which can be used to change position of the dot level.
This factor leads to the cancelation of $\Delta\eps_{\rm exch} ^ {\rm P/AP}$ at the particle-hole symmetry point of the model,
i.e. $\eps_d=-U/2$, irrespective of magnetic configuration and asymmetry.
The second factor, $\sim \beta ^{\rm P/AP}$, on the other hand, is associated with asymmetry in the system
and depends on magnetic configuration, see Eqs.(\ref{Eq:bP})-(\ref{Eq:bAP}).
It may either increase or suppress the exchange field, depending on the magnetic configuration
and parameters of the model.

As can be seen from Eq.~(\ref{Eq:Glin}), the main quantity to be calculated
is the spin-resolved spectral function of the dot-level.
This is performed with the aid of the numerical renormalization group method
with full density matrix (fDM-NRG). This method, known as the most powerful and versatile to 
study various quantum impurity problems, allows us to determine the dependence
of the spectral function on parameters of the system in most accurate and reliable way.
The starting point for the NRG is logarithmic discretization of the conduction 
band of the leads and mapping of the initial Hamiltonian to the Hamiltonian 
of a tight-binding chain with exponentially decaying hoppings, the so-call Wilson chain~\cite{WilsonRMP75}.
The chain Hamiltonian is then solved in an iterative manner and its discarded eigenstates
are used for the construction of a full density matrix~\cite{WeichselbaumPRL07},
which enables the calculation of relevant static and dynamic quantities at arbitrary temperature.
In our calculations we have in particular employed the flexible
density-matrix numerical renormalization group code~\cite{FlexibleDMNRG}.
In calculations we kept $1024$ states at each iteration and used the Abelian symmetries
for the total spin $z$th component and the total charge.

\section{Results and discussion}

In the following we present and discuss the numerical results on 
the linear conductance and spin polarization of the current in both
magnetic configurations as well as the resulting TMR effect.
First, we analyze the effects related with the contact asymmetry, $\G_L \neq \G_R$, and then
the effects due to different spin polarizations of the left and right lead, $p_L\neq p_R$, are discussed.
Finally, we present the general case when both asymmetries are present.

\subsection{Effects of left-right contact asymmetry, $\G_L\neq\G_R$}

\begin{figure}[t]
  \includegraphics[width=0.6\columnwidth]{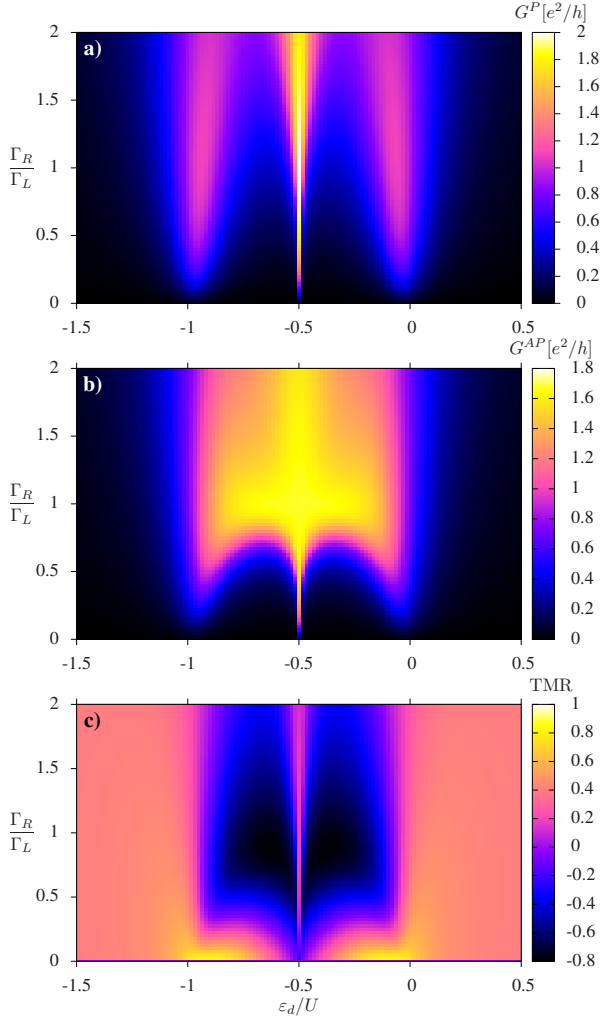}
  \caption{\label{Fig:2}
   (Color online) The zero-temperature linear conductance in the parallel (a) and antiparallel (b) magnetic configuration
   and the resulting TMR (c) as a function of the level position $\eps_d/U$
   and the left-right couplings ratio $\G_R/\G_L$. The parameters are:
   $U=0.12W$, $\G_L=0.005W$ and $p_L=p_R=0.4$, with $W\equiv 1$ the band halfwidth.
   }
\end{figure}

By changing the ratio $\G_R / \G_L$ one changes both the magnitude of exchange field
as well as the Kondo temperature. However, the dependence of both quantities on the strength
of coupling $\G=\G_L+\G_R$ is different: while $| \Delta \eps_{\rm exch}| \sim \G$, the Kondo temperature $T_K$ depends on $\G$
in an exponential way~\cite{Haldane78,Costi94}, $T_K = \sqrt{U\G/2}\; \exp\left[ \pi \eps_d (\eps_d+U) / (2U\G) \right]$
(for $p_L=p_R=0$). Tuning the contact asymmetry ratio will thus change the ratio
$| \Delta\eps_{\rm exch} | / T_K$, which conditions the occurrence of the Kondo effect,
as discussed and presented in the following.

Figure~\ref{Fig:2} shows the linear conductance in the parallel and antiparallel configuration,
as well as the resulting TMR effect as a function of the couplings' ratio $\G_R/\G_L$ and the level position
$\eps_d / U$. Experimentally, the position of the dot level can be changed by
tuning the gate voltage. When $\G_R / \G_L = 0$, the dot is coupled only to a single (left) lead and the conductance
through the system is obviously equal to zero. With increasing the coupling to the right lead, the 
linear conductance becomes finite and exhibits a strong dependence on the level position.
In the elastic cotunneling regime, i.e. for $\eps_d / \G \gg 0$ or $(\eps_d + U) / \G \ll 0$,
the conductance in the parallel configuration is $G^{\rm P} \sim (1+p_Lp_R)\G_L \G_R$,
while for the antiparallel configuration one gets, $G^{\rm AP} \sim (1-p_Lp_R)\G_L \G_R$, 
yielding ${\rm TMR} = 2p_Lp_R / (1-p_L p_R)$~\cite{weymannPRB11,weymannPRB05},
which for parameters assumed in Fig.~\ref{Fig:2} gives, ${\rm TMR} \approx 0.38$.
Note that this value is independent of the couplings asymmetry, provided
the current is mediated by elastic cotunneling events.

\begin{figure}[t]
  \includegraphics[width=0.95\columnwidth]{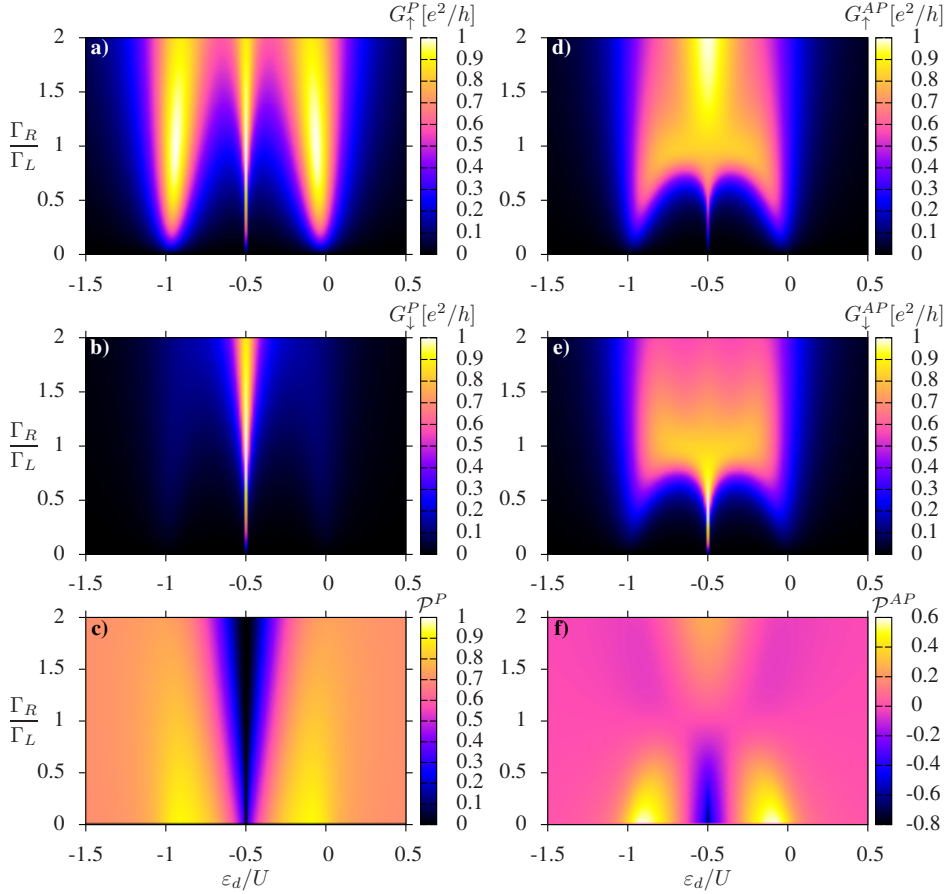}
  \caption{\label{Fig:3}
   (Color online) The spin-resolved linear conductance in the parallel (a,b) and antiparallel (d,e) magnetic configuration
   for the spin-up (a,d) and spin-down (b,e) channels,
   and spin polarization of the current in the parallel (c) and antiparallel (f) configuration
   as a function of $\eps_d/U$ and $\G_R/\G_L$. The parameters are the same as in Fig.~\ref{Fig:2}.
   }
\end{figure}

In the Coulomb blockade regime, on the other hand, the situation is more complex,
since the dot is singly occupied and the electronic correlations may lead to the Kondo effect.
Moreover, the occurrence of the Kondo effect is conditioned by the ratio of the Kondo temperature
and the exchange field induced splitting. In the parallel configuration, the exchange field is always present, 
irrespective of $\G_R/\G_L$, and the Kondo resonance is suppressed, except for the particle-hole
symmetry point, $\eps_d = -U/2$. 
By moving away from this point, the conductance drops
once $| \Delta\eps^{\rm P}_{\rm exch} | \gtrsim T_K$. Since $T_K$ depends on the coupling strength $\G$,
increasing $\G_R / \G_L$ raises the Kondo temperature, which leads to 
larger width of the Kondo peak in the middle of the Coulomb blockade regime, see Fig.~\ref{Fig:2}(a).
In the antiparallel configuration, the exchange field vanishes if the system is symmetric, 
and there is a broad Kondo resonance in the whole local moment regime, see Fig.~\ref{Fig:2}(b)
for $\G_R / \G_L \approx 1$. Nevertheless, if $\G_R / \G_L \neq 1$, the Kondo effect becomes suppressed
when $| \Delta\eps^{\rm AP}_{\rm exch}| \gtrsim T_K$. This happens faster for $\G_R / \G_L < 1$ than for 
$\G_R / \G_L > 1$, since for smaller coupling $\G$ the Kondo temperature is lower and the above condition
can be fulfilled for relatively small asymmetries of the couplings.
As already mentioned, for the particle-hole symmetry point the exchange field vanishes
in both magnetic configurations, see Eq.~(\ref{Eq:Exch}), the zero-temperature conductance is then just given by,
$G^{\rm P/AP} = e^2/h \sum_\s 4\G_{L\s} \G_{R\s} / (\G_{L\s} + \G_{R\s})^2$, with 
the couplings correspondingly dependent on magnetic configuration of the device.
For symmetric couplings, $\G_R / \G_L  = 1$, both $G^{\rm P}$ and $G^{\rm AP}$ reach 
the maximum, with $G^{\rm P} = 2e^2/h$ and $G^{\rm AP} = (1-p_Lp_R)2e^2/h$.
The respective behavior of conductance in both magnetic configurations
leads to the corresponding dependence of the TMR, which is shown in Fig.~\ref{Fig:2}(c).
Generally, the TMR is negative in the whole blockade regime, except for the particle-hole symmetry point,
which is associated with the fact that $|\Delta\eps^{\rm P}_{\rm exch}| > |\Delta\eps^{\rm AP}_{\rm exch}|$,
and consequently $G^{\rm P} < G^{\rm AP}$. Only for $\eps_d = -U/2$, when the exchange field is suppressed,
one finds a typical spin-valve effect with positive tunnel magnetoresistance.
Thus, tuning the position of the dot level and the asymmetry factors,
one can obtain a device with desired magnetoresistive properties.

Another quantity describing the spin-resolved transport properties of the system,
interesting from an application point of view, is the spin polarization
$\mathcal{P}^{\rm P/AP}$ of the current flowing through the device,
which is shown in Figs.~\ref{Fig:3}(c) and (f) for both magnetic configurations.
The behavior of spin polarization can be understood from the spin-resolved conductance.
In the parallel magnetic configuration the coupling of the spin-up level is much
stronger than the coupling of the spin-down level, since spin-up electrons
belong to the spin-majority band. As a consequence,
the spin-up channel gives the main contribution to the conductance, 
see Figs.~\ref{Fig:3}(a) and (b). The difference between $G^{\rm P}_\up$
and $G^{\rm P}_\down$ is most visible around the resonances, $\eps_d\approx 0$ and $
\eps_d\approx -U$, where the spin polarization $\mathcal{P}^{\rm P}$ takes large positive values,
approaching unity for $\G_R/\G_L \ll 1$. This is contrary to the region around the particle-hole
symmetry point where $G^{\rm P}_\up \approx G^{\rm P}_\down$  and
the spin polarization is suppressed, $\mathcal{P}^{\rm P} \to 0$.
In the antiparallel configuration the situation is slightly more complex,
since the spin-resolved couplings depend greatly on the system's asymmetry, see Eq.~(\ref{Eq:Gammaeff}).
For equal spin polarizations of the leads, $p_L=p_R$, as assumed in Fig.~\ref{Fig:3},
one finds the effective couplings, $\G_\s^{\rm AP} = \G + \s p (\G_L-\G_R)$.
In consequence, the spin-resolved couplings fulfill the following relations,
$\G_\up^{\rm AP} > \G_\down^{\rm AP}$ for $\G_R/\G_L<1$,
and $\G_\up^{\rm AP} < \G_\down^{\rm AP}$ for $\G_R/\G_L>1$, with 
$\G_\up^{\rm AP} =\G_\down^{\rm AP}$ for the symmetric case, $\G_L=\G_R$.
This gives rise to the corresponding behavior of the spin polarization:
$\mathcal{P}^{\rm AP} \lessgtr 0$ for $\G_R/\G_L \lessgtr 1$,
and $\mathcal{P}^{\rm AP} \approx 0$ for $\G_R/\G_L \approx 1$.
In the region around $\eps_d = -U/2$, where the effective field vanishes, the behavior of 
$\mathcal{P}^{\rm AP}$ is however different.
The spin-dependent conductances are then given by,
$G^{\rm AP}_\s = 4 e^2/h \, (1-p^2)\G_L\G_R / [\G +\sigma p (\G_L - \G_R)]^2$, which yields
negative spin polarization $\mathcal{P}^{\rm AP}<0$ for $\G_R / \G_L <1$, and 
positive spin polarization $\mathcal{P}^{\rm AP}>0$ for $\G_R / \G_L >1$.
Note that this is just opposite to the case when the exchange field is present,
i.e. $\eps_d \neq -U/2$.
The above analysis clearly demonstrates that by properly engineering the couplings
between the dot end electrodes, and by tuning the occupancy of the dot with the gate voltage,
one can obtain desired spin polarizations of the flowing current,
spanning almost the whole range from $-1$ to $1$.

\begin{figure}[t]
  \includegraphics[width=0.65\columnwidth]{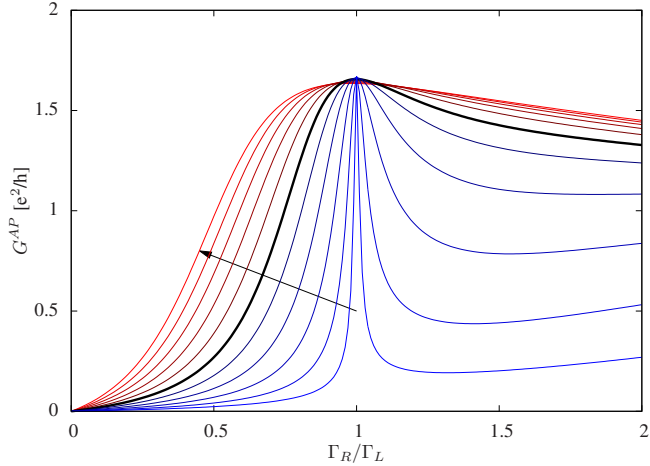}
  \caption{\label{Fig:4}
   (Color online) The linear conductance in the antiparallel configuration
   as a function of left-right contact asymmetry $\G_R/\G_L$
   for different values of the coupling to the left lead $\G_L$ for $p_L=p_R=0.4$.
   Each line corresponds to different value of $\Gamma_L$, increasing
   in the direction indicated by the arrow
   from $\G_L=0.0025W$ to $\G_L=0.0075W$ in step of $0.0005W$.
   The thick black line corresponds to $\Gamma_L = 0.005W$, value used in previous figures.
   The parameters are the same as in Fig.~\ref{Fig:2} with $\eps_d = -U/3$.
   }
\end{figure}

From an experimental point of view, it may be important to know how large the asymmetry should be
to suppress the Kondo resonance in the antiparallel configuration.
To address this question, in Fig.~\ref{Fig:4} we plot the linear conductance
as a function of $\G_R / \G_L$ for different values of the coupling $\G_L$.
Since $T_K$ depends exponentially on $\G$, 
changing $\G_L$ corresponds to a huge change in the Kondo temperature (note that $T_K$ also depends on $\G_R / \G_L$).
On the other hand, the dependence of the exchange field on $\G_L$ and the ratio $\G_R / \G_L$ is only algebraic.
It can be seen that with lowering $\G_L$, the suppression of the Kondo effect occurs for smaller
asymmetries, e.g. for very weak coupling even relatively small asymmetry between the left-right contacts
can fully suppress the linear conductance in the antiparallel configuration.
Proper and very careful implementation of a quantum dot/molecular device is therefore necessery
in order to observe desired effects, such as e.g. restoration of the Kondo effect
when switching the magnetic configuration from parallel into antiparallel one~\cite{pasupathy_04}.

\subsection{Effects of different spin polarizations of the leads, $p_L\neq p_R$}

\begin{figure}[t]
  \includegraphics[width=0.6\columnwidth]{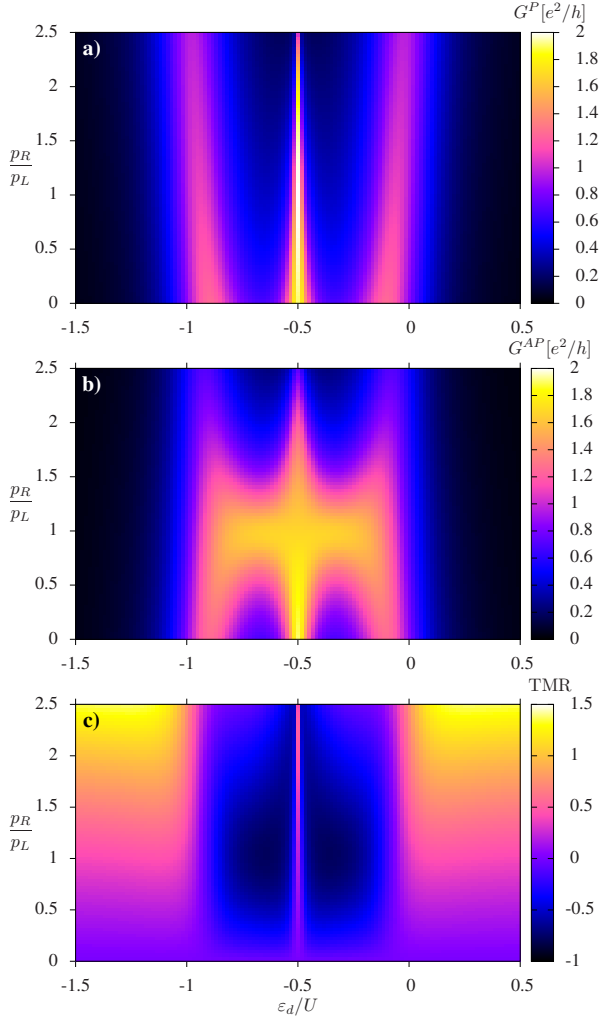}
  \caption{\label{Fig:5}
   (Color online) The conductance in the parallel (a) and antiparallel (b) configuration
   and the resulting TMR (c) as a function of $\eps_d/U$
   and the leads' spin polarization ratio $p_R/p_L$. The parameters are
   the same as in Fig.~\ref{Fig:2} with $\G_L = \G_R = 0.005W$ and $p_L=0.4$.}
\end{figure}

Up to now we have focused on the asymmetry related to the left-right contacts,
however, the asymmetry can be also present if the electrodes have different spin polarizations, $p_L\neq p_R$.
The corresponding transport characteristics are shown in Fig.~\ref{Fig:5}
for equal couplings $\G_L=\G_R$ and different spin polarization ratio $p_R/p_L$, with $p_L=0.4$.
The maximum value, $p_R/p_L = 2.5$, corresponds then to fully spin-polarized
right lead, while $p_R/p_L = 0$ corresponds to nonmagnetic right electrode.
Now, by changing $p_R/p_L$, one can tune the
magnitude of the exchange field, while the relevant Kondo temperature is constant.
Generally, with increasing the ratio $p_R/p_L$,
the average spin polarization increases, and so does the exchange field.
This intuitive behavior is however only valid for the parallel magnetic configuration
and is nicely visible in Fig.~\ref{Fig:5}(a).
It can be seen that $G^{\rm P}$ displays a Kondo resonance at the particle-hole symmetry point,
whose width decreases with increasing strength of the exchange field,
i.e. with increasing the ratio $p_R/p_L$.
In the antiparallel configuration, on the other hand, the exchange field is a nonmonotonic function
of $p_R/p_L$: it is maximum for $p_R =0$, vanishes for $p_R = p_L$,
and again reaches local maximum for $p_R = 1$.
Consequently, $G^{\rm AP}$ displays the Kondo effect in the whole
Coulomb blockade regime when $p_L \approx p_R$, which becomes then suppressed with changing the 
materials' asymmetry ratio from the point $p_R/p_L=1$, see Fig.~\ref{Fig:5}(b).
Moreover, it can be seen that the Kondo resonance around $\eps_d = -U/2$
is now broader than in the case of parallel configuration since generally,
$|\Delta \eps_{\rm exch}^{\rm P} |> |\Delta \eps_{\rm exch}^{\rm AP}|$.
The different dependence of $G^{\rm P}$ and $G^{\rm AP}$ on the spin polarization ratio $p_R/p_L$
reflects itself in a nontrivial behavior of the TMR effect, see Fig.~\ref{Fig:5}(c).
The TMR is positive in the whole elastic cotunneling regime and given by the Julliere value,
while it takes large negative values in the local moment regime for $p_R/p_L\approx 1$ when ${\rm TMR}\to -1$,
and becomes again positive in the middle of the Coulomb blockade regime.

\begin{figure}[t]
  \includegraphics[width=0.6\columnwidth]{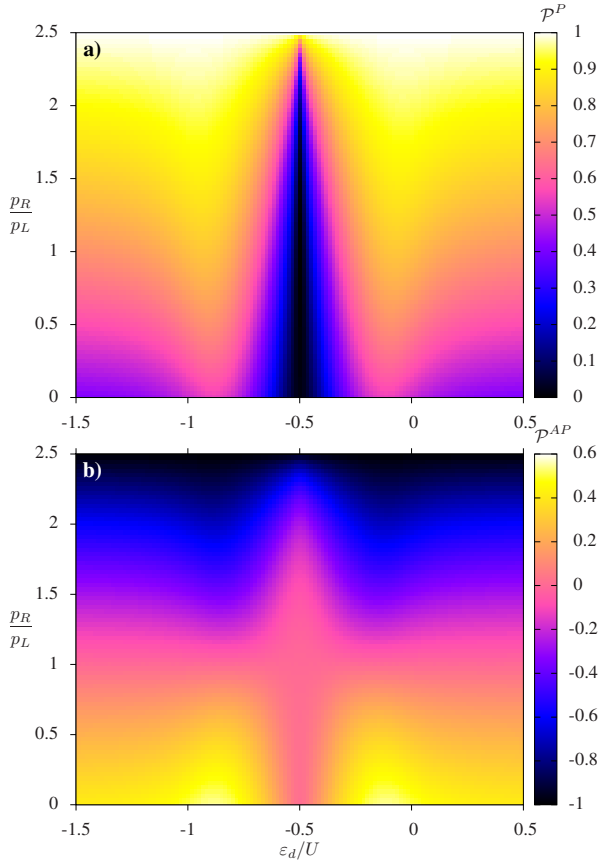}
  \caption{\label{Fig:6}
   (Color online) The spin polarization of the current in the parallel (a) and antiparallel (b)
   magnetic configuration as a function of $\eps_d/U$ and $p_R/p_L$
   for parameters as in Fig.~\ref{Fig:5}.
   }
\end{figure}

The spin polarization of the conductance in both magnetic configurations
as a function of level position and the leads' spin polarization ratio $p_R/p_L$
is shown in Fig.~\ref{Fig:6}. In the parallel configuration, the dependence of
$\mathcal{P}^{\rm P}$ is quite intuitive: the spin polarization increases
with raising the average spin polarization $p$, i.e. increasing $p_R/p_L$,
and reaches unity for $p_R\to 1$, since then the right electrode supports 
only spin-up electrons. Finite spin polarization is mainly due to the presence of exchange field.
Consequently, in the middle of the Coulomb blockade valley the spin polarization
is suppressed, because $\Delta\eps^{\rm P}_{\rm exch} \to 0$.
In the antiparallel configuration, the spin polarization changes sign
with increasing the ratio $p_R/p_L$. When, $p_L > p_R$, there are more spin-up
tunneling processes than spin-down ones and the spin polarization is positive,
while for $p_L < p_R$, the sign of $\mathcal{P}^{\rm AP}_{\rm exch}$ is 
determined by the majority band of the right lead (spin-down electrons),
therefore $\mathcal{P}^{\rm AP}_{\rm exch} < 0$, reaching $-1$ for $p_R\to 1$.
On the other hand, once $p_R \approx p_L$, the spin polarization vanishes
since the resultant couplings to spin subbands are comparable.

\subsection{Effects of both contact and material asymmetry, $\G_L\neq\G_R$ and $p_L\neq p_R$}

\begin{figure}[t]
  \includegraphics[width=0.6\columnwidth]{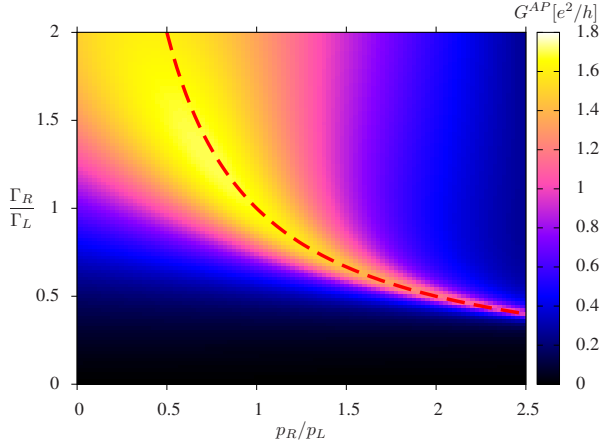}
  \caption{\label{Fig:7}
   (Color online) The linear conductance in the antiparallel configuration
   as a function of left-right contact asymmetry $\G_R/\G_L$
   and the leads' spin polarization asymmetry $p_R/p_L$. The parameters are
   the same as in Fig.~\ref{Fig:2} with $\eps_d = -U/3$.
   The dashed line corresponds to parameters for which the condition (\ref{Eq:exchvan}) is met.
   }
\end{figure}

From our discussion follows that asymmetry can destroy the Kondo resonance
in the antiparallel configuration if $|\Delta \eps_{\rm exch}^{\rm AP} | \gtrsim T_K$.
However, it turns out that even if there is an asymmetry in the system, either related to the contacts
or to the material, there is a parameter range when the exchange field can still vanish
in the antiparallel configuration. This happens precisely when the condition (\ref{Eq:exchvan})
is met, so that $\G_\up^{\rm AP} = \G_\down^{\rm AP}$.
Figure~\ref{Fig:7} shows the dependence of the linear conductance in the antiparallel magnetic configuration
as a function of the relevant asymmetry parameters, i.e. $p_R/p_L$ and $\G_R/\G_L$, for $\eps_d = -U/3$.
Away from the particle-hole symmetry point, $\eps_d \neq -U/2$, the presence and the strength of the exchange field
are conditioned by the system's asymmetry parameters, see Eq.~(\ref{Eq:Exch}).
For $p_R / p_L  \to 0$, the conductance $G^{\rm AP}$ behaves similarly as that for
the parallel configuration, since then trivially there is only one magnetic electrode.
The conductance is then generally suppressed and starts increasing only
for $\G_R/\G_L> 1$, when the Kondo temperature increases, so that the condition
$| \Delta \eps_{\rm exch}^{\rm AP}| \gtrsim T_K$ is only weakly met or even not met.
For $p_R/p_L > 0$, the case becomes much more interesting, since
for certain asymmetry parameters one finds an increased conductance due to the Kondo effect.
This happens when the condition (\ref{Eq:exchvan}) is satisfied, as
presented by a dashed line in Fig.~\ref{Fig:7}. The exchange field increases then
as one moves away from this line. 
The width of the restored Kondo resonance increases with increasing $\G_R / \G_L$,
since a larger asymmetry is then needed to suppress the Kondo effect, which happens for
$|\Delta \eps_{\rm exch}^{\rm AP}| \gtrsim T_K$.

\section{Conclusions}

In this paper we have studied the Kondo effect in quantum dots asymmetrically coupled 
to ferromagnetic leads. The calculations were performed with the aid
of the numerical renormalization group method. In particular, we have determined
the dependence of the linear conductance in different magnetic configurations of the device,
the TMR and current spin polarization on the contacts' and material's asymmetry parameters.
For quantum dots symmetrically coupled to external leads, the conductance in the parallel configuration is suppressed
due to the presence of exchange field, while in the antiparallel configuration
the exchange field is absent. On the other hand, when the dot is coupled asymmetrically
to the leads, we have shown that the exchange field can also develop in the antiparallel configuration.
If the magnitude of exchange field induced level splitting is larger than the Kondo temperature, the Kondo effect
becomes suppressed, which may happen even for relatively small asymmetry between the couplings
to the left and right lead.
This leads to a nontrivial dependence of the tunnel magnetoresistance on the asymmetry parameters.
In addition, we have demonstrated that even if the system is asymmetric,
there is a range of parameters in the antiparallel configuration where the exchange field can still vanish.
We have also derived approximate analytical formulas for the exchange field
in both magnetic configurations depending on the asymmetry parameters.

Concluding, the presented analysis shows that by properly tuning the parameters of the system,
one can in principle construct a magnetoresistive device of desired properties,
including a source of spin polarized electrons with basically any designed spin polarization.
Moreover, a very careful fabrication of a sample seems inevitable,
if one wants to observe the suppression and restoration
of the Kondo effect by changing the magnetic configuration of the device.


\ack

This work was supported by the Polish Ministry of Science and
Higher Education through a research project No. N N202 199739 in years 2010-2013
and a 'Iuventus Plus' project No. IP2011 059471 in years 2012-2014.
I.W. also acknowledges support from
the EU grant No. CIG-303 689 and the Alexander von Humboldt Foundation.


\end{document}